\documentclass[twocolumn,showpacs,preprintnumbers,amsmath,amssymb,prb]{revtex4}

\usepackage{graphicx}
\usepackage{dcolumn}
\usepackage{hyperref}
\usepackage{bm}

\begin{document}

\title{Depth dependent dynamics in the hydration shell of a protein}

\author{J. Servantie}
\author{C. Atilgan}
\author{A.~R. Atilgan}

\affiliation{Faculty of Engineering and Natural Sciences, 
Sabanci University, Orhanli 34956 Tuzla, Istanbul, Turkey}

\begin{abstract}
We study the dynamics of hydration water/protein association in
folded proteins, using lysozyme and myoglobin as examples. Extensive
molecular dynamics simulations are performed to identify underlying
mechanisms of the dynamical transition that corresponds to the onset
of amplified atomic fluctuations in proteins. The number of water
molecules within a cutoff distance of each residue scales linearly
with protein depth index and is not affected by the local dynamics
of the backbone. Keeping track of the water molecules within the
cutoff sphere, we observe an effective residence time, scaling
inversely with depth index at physiological temperatures while the
diffusive escape is highly reduced below the transition. A depth
independent orientational memory loss is obtained for the average
dipole vector of the water molecules within the sphere when the
protein is functional. While below the transition temperature, the
solvent is in a glassy state, acting as a solid crust around the
protein, inhibiting any large scale conformational fluctuations. At
the transition, most of the hydration shell unfreezes and water
molecules collectively make the protein more flexible.
\end{abstract}

\maketitle

\section{INTRODUCTION}

Water is essential for the stability and function of proteins. It
modifies properties of proteins such as the display of the dynamic
transition temperature, $T_d$. The dynamical transition corresponds 
to the glass transition observed in the hydration shell of a protein accompanied 
by a correlated rubber-glass transition of the protein. The transition results 
in enhanced atomic fluctuations with respect to dehydrated proteins and consequentely 
increased flexibility of the backbone. \cite{Doster1989}
The dynamical transition temperature depends on the time scale of the 
observation. Since the observed configurational fluctuations are delimited by 
the resolution of the measurement, the reported value of the dynamical transition 
temperature depends on that resolution. \cite{Doster2008} For example, neutron 
scattering data yields $T_d=240$ K at a time window of 50 ps while it decreases 
to $T_d=220$ K at a time scale of 2 ns. \cite{Doster2010a,Doster2010b} The transition 
is accompanied by an increase in the slope of the 
mean squared atomic fluctuations. The local relaxation times of
$C_\alpha$ atoms is observed to change in accord with the increased flexibility 
of the backbone. \cite{Atilgan2008} However, in the absence of solvent, the 
dynamical transition is not observed, \cite{Tsai2001} and to recover the 
transition the charged residues of the protein have to be covered by 
solvent. \cite{Steinbach1996,Leitner2003} 

Proteins in turn affect the dynamics of water in the
hydration shell. Experiments at room temperature show that the
relaxation times of water molecules in the vicinity of proteins are
larger than in the bulk \cite{Doster2005} and highly dependent on
the location around the protein. \cite{Zhong2007} While coupling of 
water dynamics with the onset of the dynamical transition has been 
investigated, \cite{Arcangeli1998,Vitkup2000,Tarek2002} and it is widely 
accepted that the solvent dictates the transition, the mechanism by which 
water operates on proteins is still an open question. \cite{Kumar2006} Recently, 
single molecule rotational correlation time of water molecules around the 
protein has been determined from spin relaxation experiments at various temperatures. 
This is more localized than translational diffusion, and in turn is 
a more accurate measure of mobility in the hydration layer. \cite{Halle2009} 
Since it is crucial to have a better understanding of the dynamics of water 
and its effect on the protein both below and above $T_d$, we study the 
translational and rotational motion of water in the vicinity of the protein 
in these separate regions, along with that of pure water and the protein. 
This study enables us to understand water dynamics on the time scales that are relevant to
protein motions.

The paper is organized as follows: In Sec. \ref{SECNUMMETHOD}, we describe 
the numerical methods and the details of the molecular dynamics (MD) 
simulations carried out for pure water and proteins in water. In 
Sec. \ref{SECRESULTS}, we first determine the glass 
transition temperature of pure water, described by the TIP3P model. 
Thereafter, we investigate the local dynamics of two commonly studied 
proteins, lysozyme and myoglobin, at room temperature as well as at 
180 K. The latter is below the dynamical transition temperature of common 
proteins, but above the glass transition of water. The hydration levels 
and water residence times around the residues are computed, and the 
predominant contributions to the observed 
behavior are shown to be accounted for by two simple models, developed 
in this work. Afterwards, the $C_\alpha$ and water relaxation times and 
their temperature dependence are investigated to uncover the nature of the 
coupling between local protein dynamics and water. Finally, conclusions 
are drawn in Sec. \ref{SECCONC}.

\section{NUMERICAL METHODS}
\label{SECNUMMETHOD}

To study the dynamics of water in the hydration shell of a protein,
we carried out MD simulations for two different proteins, lysozyme 
of 129 residues and apomyoglobin of 152 residues (Protein Data Bank 
codes 6lzm and 1jp6, respectively). Observing common patterns of a 
hydration shell for these two proteins, one may deduce universal 
properties of hydration water dynamics. The proteins are solvated 
so that the water to protein mass ratio is $h=3.42$ and $h=3.34$ 
for lysozyme and myoglobin, respectively. These values are weþþ 
above the minimum of $h=0.7$ for which a fully hydrated protein 
is observed, e.g., for bovine pancreatic trypsin inhibitor. \cite{Leitner2003}
 
The solvated simulation boxes are equilibrated for 2 ns at 300 K, 
and then a further 2 ns at the desired temperature. All simulations 
are in the NPT ensemble at a pressure of 1 atm. Two sets of simulations 
are carried out, one at 180 K, below the dynamical transition temperature, 
and the other at 300 K. The NAMD package \cite{NAMD} is used with the 
CHARMM 27 \cite{CHARMM} force field and the TIP3P water model. A time 
step of 2 fs is used and 50 ns long trajectories are
produced. Coordinates are recorded every 0.4 ps. Further details on
the simulation methods are as in Okan {\it et al}. \cite{Atilgan2009}

We investigate the equilibrium characteristics and the dynamics of
water molecules in the vicinity of each residue. The hydration shell
around a given residue is defined to contain the water molecules
within $6 \AA$ of its $C_\alpha$ atom. Smaller values of the cutoff
results in a too small number of solvent molecules, and hence poor
statistics, while larger cutoff radius will result in taking into
account more of the bulk dynamics. We note that the correlations in the 
solvent can extend up to a distance of $10 \AA$ away from the protein 
surface. \cite{Makarov1998,Ebbinghaus2007}

To differentiate the relative importance of the bulk versus the local
protein environment, we display the characteristic properties of the
residues as a function of the depth index, $d$. \cite{Varrazzo2005} This index 
is defined as twice the ratio of the accessible volume of an atom $V_a$ to 
the exposed volume of an isolated atom $V_0$,
\begin{equation}
d=\frac{2 V_a}{V_0}.
\label{EQDEPTH}
\end{equation}
Thus, the deeper the residue is in the protein, the smaller is 
the depth index. A surface residue has approximately half the accessible 
volume of an isolated atom, and its depth index is 
close to one. For an isolated atom in bulk water, the depth index is two. 
The exposed volume $V_a$ is calculated in a sphere of radius $r$. If the 
radius is chosen too small or too large, all values of depth indices converge 
to 0 or 2. The optimal value of $r$ corresponds to the case when only one residue 
has a vanishing depth index; for proteins in this work one finds $r=9\AA$. In 
all the results presented as a function of depth index, we average over residues 
with similar depth indices, where the bin sizes are evaluated from $d_{\rm max}/15$. 
The resulting error bars are shown on the figures.

To evaluate the temperature dependence of the dynamics we
also perform MD simulations of TIP3P bulk water, and hydrated
lysozyme for a series of temperatures. For the former, we span the
temperature range of 80 - 350 K at 1 atm, for 10 ns on a system of
273 water molecules. For the latter, we perform the MD simulations in
the range 160 K to 300 K, each of length 24 ns, keeping all other
conditions the same as the protein-water systems described above.

\section{RESULTS AND DISCUSSION}
\label{SECRESULTS}

\subsection{Hyration levels and water residence times}
\label{SECHYDLEVEL}

One can measure the hydration levels of each residue, defined as the average 
number of water molecules within $6 \AA$ of its $C_\alpha$ atom present. The number 
of water molecules in the hydration
shell of a residue is highly dependent on depth: The deeper
inside the protein, the higher the coordination number of a residue
is and the lower its accessible volume. Although the solvent density 
$\rho$ depends on depth, i.e. the first layer around the protein has 
a density about 15\% larger than bulk water, \cite{Merzel2002} we assume $\rho$ 
does not depend on depth index as a first approximation. The number of water 
molecules $N_w$ is then proportional to the accessible volume, $V_0\; d/2$,
\begin{equation}
N_w=\frac{\rho\; V_0}{2}\; d\;. \label{EQNUMA}
\end{equation}
Since the bulk density $\rho$ is known, one may predict the number
of water molecules as a function of the depth index using this relationship.
\begin{figure}[h!]
\centering
\includegraphics[angle=270,scale=0.55]{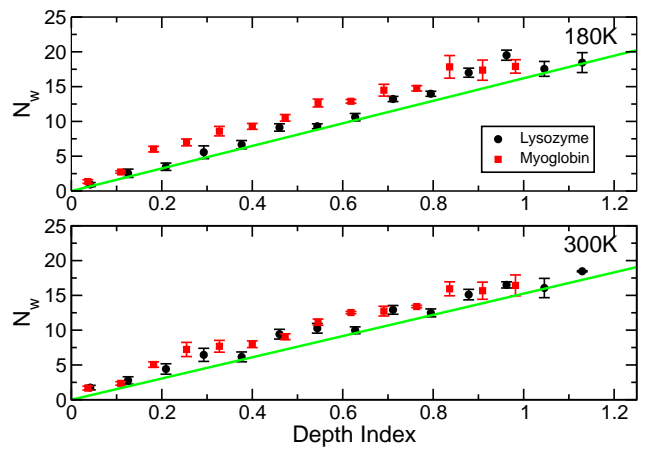}
\caption{Number of water molecules as a function of the depth index at
180 K and 300 K for lysozyme and myoglobin. The straight lines are obtained
from Eq. \ref{EQNUMA}.}
\label{FIG2}
\end{figure}
We compute the average number of water molecules both in the
hydration shell of each residue and in the bulk. The bulk values are
32.4 and 30.5 water molecules at 180 K and 300 K, respectively,
corresponding to a density of 1070 ${\rm kg/m}^3$ and 1007 ${\rm
kg/m}^3$. The results are depicted in Fig. \ref{FIG2}, and are in
agreement with the linear dependence predicted by Eq. \ref{EQNUMA} for both proteins. Hence, 
the temperature dependence of the distribution of water molecules in the protein is 
determined predominantly from the bulk value $\rho(T)$. Local structural irregularities 
and protein dynamics have a relatively minor role.

\begin{figure}
\centering
\includegraphics[angle=270,scale=0.55]{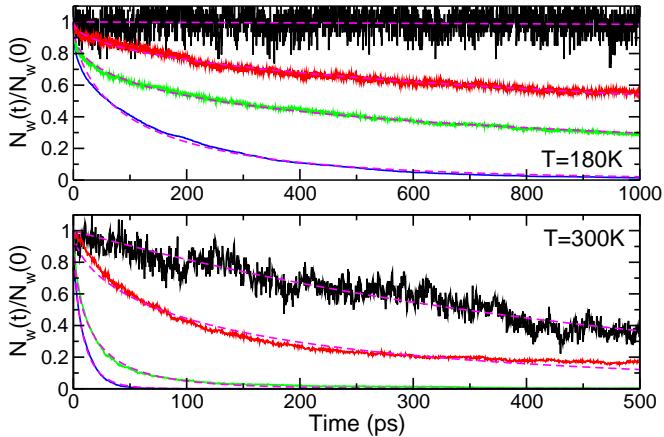}
\caption{Normalized number of water molecules initially residing in the
vicinity of a residue as a function of time. Samples are for
residues with different depth indices (CYS30: $d=0.065$, ASN65:
$d=0.435$, THR47: $d=1.26$) and in the bulk ($d=2$). The dashed
lines represent the stretched exponential fits. $d$ increases from
the top towards the bottom curves.} 
\label{FIGnwt}
\end{figure}

\begin{figure}
\centering
\includegraphics[angle=270,scale=0.55]{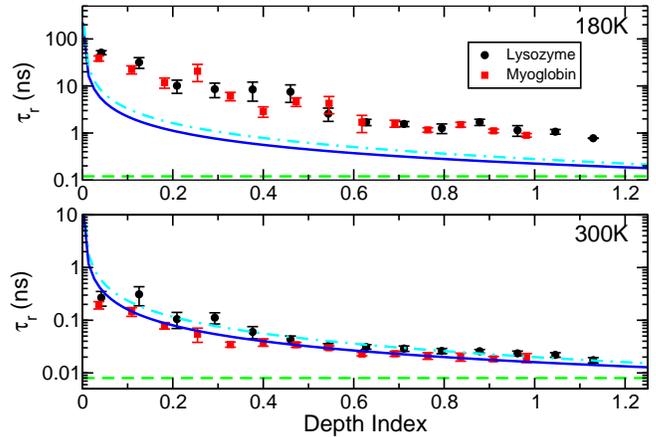}
\caption{Residence times of water molecules as a function of the
depth index for lysozyme and myoglobin from fits to Eq. \ref{STRFIT}. 
The stretched exponents fluctuate around 0.5 at all
depths and temperatures for both proteins. The limiting value in the
bulk is represented by the dashed lines. Predictions from Eq. \ref{EQRESTIME} 
(solid line) 
and from Bezrukov {\it el al.} \cite{Bezrukov2001} (dot-dashed line) 
are also shown.} 
\label{FIG3}
\end{figure}

Although the distribution of water molecules, an equilibrium
property, is not affected by temperature, the dynamics of water
around the protein depends strongly on $T$. We characterize the
latter by their residence times, $\tau_r$, which is the average time
a molecule takes to escape from a given region. To measure the
residence times of water molecules in the hydration shells of all
residues, we partition the 50 ns long trajectories into fifty 1 ns
long pieces, and record the initial number of hydration shell water
molecules in each region. We then compute the decrease in the number
of these water molecules as a function of time by checking if they remain 
in the hydration shell. Since the trajectories are recorded in 
intervals of 0.4 ps, water molecules that cross the boundary and 
return to the monitored layer faster than that time scale are assumed 
to have remained. We depict in Fig. \ref{FIGnwt} this decrease for selected 
residues of lysozyme, a highly buried one CYS30 ($d=0.065$), one at an 
intermediate depth ASN65 ($d=0.435$), a surface residue THR47 ($d=1.26$) 
and finally the escape process in the bulk ($d=2$).
Many different processes contribute to the curves in Fig. \ref{FIGnwt}. Most 
of the water molecules are close to the edge of the sphere we consider, and 
thus can escape faster on average. On the other hand, the water molecules at 
the center of the sphere initially will remain longer. There will also 
be additional contributions from the fluctuating chain molecule. Consequently, 
one expects to have a superposition of exponential decays for $N_w(t)$ which 
is best described by a stretched exponential function, \cite{Watts1970}
\begin{equation}
N_w(t)=\langle N_w \rangle\; \exp{\left[-(t/\tau_r)^\beta\right]},
\label{STRFIT}
\end{equation}
where $\tau_r$ is the effective residence time and $\beta$ is the
stretched exponent. The fits are illustrated in Fig. \ref{FIGnwt} and are in 
good agreement with the MD results, validating the model. The
stretched exponent has a value $\beta\approx0.5$ regardless of the
location of the residue and temperature. We thus compute the residence times 
for all the residues of lysozyme and myoglobin, and plot them as a function 
of depth index in Fig. \ref{FIG3}.
The residence times for equivalent depths can have important
variations because of the residue specificities and their spatial
arrangements. For example, residues can be hydrophobic or
hydrophilic, and consequently decrease or increase the residence
time of the surrounding water. Moreover, residues can be
concentrated in one region of the hydration shell facilitating the
escape process. However, one can capture the general trend when
considering a large number of residues. Differences of up to
three orders of magnitude between the residence times of solvent
molecules buried in the protein compared to those near the surface
are observed. This is due to the fact that the water molecules
inside have to escape traps formed by the residues. Thus, the higher
the coordination number of a residue, the larger the number of
traps, and the longer the escape time. Note that, for certain residues 
that display extremely slow decay profiles at 180 K (exemplified by the 
uppermost curve in Fig. \ref{FIGnwt}), Eq. \ref{STRFIT} provides residence 
times longer than 100 ns. This is due to the error involved in extrapolation 
of the initial decay profiles to very long time scales.

Using a model of diffusive motion in the presence of local
obstacles, a relationship may be derived for the residence time as a
function of depth. As a first order approximation, we assume the
probability to escape from the neighborhood of a residue $P_r$
decreases linearly with the occupied volume as
\begin{equation}
P_r=P_0 \left(1-\frac{V_r}{V_0}\right)\;,
\end{equation}
where $P_0$ is the probability of escaping in the bulk and
$V_r=\sum_{j=1}^{N_c} V_j$ the occupied volume around the residue.
The occupied volume is the total volume minus the available volume
$V_r=V_0-V_a$; using $d=2V_a/V_0$ (Eq. \ref{EQDEPTH}), one gets
$P_r=P_0\; d/2$. Since the residence time scales inversely with the
probability of escape, \cite{Gaspard1998} and the proportionality
constant is assumed
to be an intrinsic property of the escaping species irrespective of
the local environment, one gets,
\begin{equation}
\tau_r(d)=\tau_B\; \frac{2}{d}\;.
\label{EQRESTIME}
\end{equation}
where $\tau_B$ is the residence time in the bulk. In Fig. \ref{FIG3}, we 
display the findings from Eq. \ref{EQRESTIME} along with the
results from the highly elaborate analytical solution of Bezrukov 
{\it et al.} \cite{Bezrukov2001} for the probability of diffusion 
displacement of a molecule using the fraction of possible trajectories 
in the presence of obstacles. The rather simple model of Eq. \ref{EQRESTIME} 
captures the main features of this detailed
analytical solution. We further find that both approaches predict
the depth dependence at 300 K, while they fail to do so at 180 K.

So far as the predictions of Eq. \ref{EQRESTIME} reproduce the
main features of water behavior around the protein at room
temperature, they corroborate that specificity plays a lesser role
in determining residence times at physiological temperatures. 
\cite{Luise2000,Makarov2000} Below $T_d$, however, we find that the
simple trapped diffusion idea does not capture the dynamics of
water. In fact, the calculated residence times are approximately 
one order of magnitude slower than predicted by the theory, indicating 
that the dynamics is not controlled by diffusion in the presence of
obstacles,  hence suggesting that hydration shell water is not liquid 
below the transition temperature. This confirms the observation that 
translational diffusion is hindered below the dynamical transition. \cite{Tournier2003}

\subsection{Depth dependent relaxation times in the hydration layer}
\label{SECRELAX}

To further investigate the water/protein interactions, we quantify
the dynamics of the protein, hydration layer, and bulk solvent by
measuring relaxation times. For the $C_\alpha$ atom of the {\it ith}
residue, we follow the decay of the correlation function,
\begin{equation}
C^\alpha_i(t)=\frac{\langle\Delta \mathbf{R}_i(t)\cdot
\Delta \mathbf{R}_i(0)\rangle}{\langle\Delta \mathbf{R}_i^2\rangle}
\label{EQCORRELCA}
\end{equation} 
where $\Delta\mathbf{R}_i(t)=\mathbf{R}_i(t)-\langle\mathbf{R}_i\rangle$ 
is the fluctuation vector of the atom. We note that $\langle \mathbf{R}_i \rangle$
is computed for each 0.5 ns trajectory piece and then a best fit
superposition is performed to separate the internal motion of the
protein from long time tumbling. Previously, this method permitted
evaluating the dynamical transition temperature for different
protein-solvent systems. \cite{Atilgan2005,Atilgan2008}

\begin{figure}
\centering
\includegraphics[angle=270,scale=0.55]{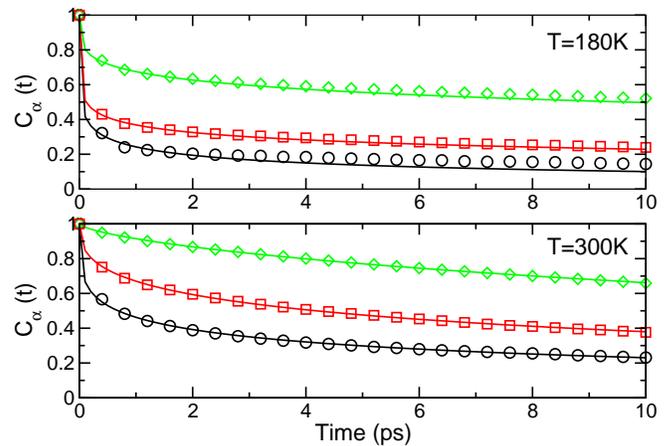}
\caption{$C_\alpha$ correlation function for residues with different
depth indices. The circles are for the buried residue (CYS30, $d=0.065$),
the squares for the intermediate (ASN65, $d=0.435$) and the diamonds
for the surface one (THR47, $d=1.26$). The straight lines represent
the stretched exponential fits.}
\label{FIGCa}
\end{figure}

Many different processes over a wide range of time scales contribute
to the backbone dynamics as we recently discussed in detail. \cite{Atilgan2009} 
Here we employ stretched exponential functions on
the correlation functions of Eq. \ref{EQCORRELCA} to fit the
initial decay processes as in our previous work. \cite{Atilgan2008} 
We depict in Fig. \ref{FIGCa} the correlation function
for the three residues of lysozyme also exemplified in Fig. \ref{FIGnwt} 
and the corresponding stretched exponential fits. We
note that the long time tails of these relaxation functions belong
to more collective dynamics with a nanosecond time scale, which we
do not further elaborate upon since this was discussed in great
detail recently. \cite{Atilgan2009}
\begin{figure}
\centering
\includegraphics[scale=0.75]{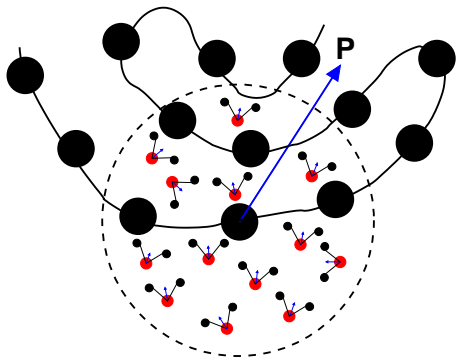}
\caption{Schematic representation of the hydration shell
of a residue and the local polarization vector $P$.}
\label{FIG1}
\end{figure}
Closely associated with backbone dynamics, we further compute the
water relaxation times by using the local polarization vector $\mathbf{P}_i$ 
in the vicinity of each $C_\alpha$ as depicted in Fig. \ref{FIG1}. One 
can then compute the following correlation function,
\begin{equation}
C^w_i(t)=\frac{\langle \Delta P_i(t) \Delta P_i(0)\rangle}
{\langle \Delta P_i^2 \rangle}
\label{EQCORRELW}
\end{equation} 
where $\Delta P_i(t)=|\mathbf{P}_i(t)|-\langle|\mathbf{P}_i(t)|\rangle$ is the
fluctuation of the norm of the polarization vector of the {\it ith}
residue. The polarization vector is the sum of the polarization
vectors of the water molecules within the hydration shell of the
residue considered,
\begin{equation}
\mathbf{P}_i(t)=\sum_{j \in i} \mathbf{p}_j(t).
\end{equation}
We emphasize that we use a difference definition utilizing the 
norms (\ref{EQCORRELW}), as opposed to the usual expression involving 
the scalar product of the dipolar moments. While the dipolar moment 
in the bulk is equal to zero on average, it is not the case inside 
the protein, where water molecules can have preferred orientations 
because of the proximity of protein atoms. Consequently we use the 
quantity $\Delta P_i(t)$ which eliminates this bias, in addition 
to the advantage of Eq. (\ref{EQCORRELW}) providing a faster decay 
since it corresponds to a higher moment. Moreover, we note that the 
relaxation time obtained depends on the number of water molecules 
considered for either definition of the relaxation.\cite{Spoel1998} 
To recover the experimental value of the relaxation time, one should 
compute the polarization vector for larger spheres. At room temperature, 
we find that the radius should be larger than the correlation length 
of the solvent which 
is ca. $12\; \AA$. Beyond this sphere size, the Debye relaxation time 
converges to the value of 7.3 $\pm$ 0.7 ps, as also reported in literature 
previously for TIP3P water model \cite{Steinhauser1998}, and compares well 
with the experimental value of 8.2 ps. In contrast, for the sphere size 
chosen in this work, $6\; \AA$, the value is 5 ps. For the relaxation 
function of Eq. (\ref{EQCORRELW}), the respective relaxation times 
are 3.2 and 1.8 ps for spheres of radius $12\; \AA$ and $6\; \AA$, 
respectively. Thus, our computed values of the relaxations are 
consistent within this work, and are useful for comparing the 
local dynamics of water clusters at various depths. However, they 
cannot be compared directly with Debye relaxation times.
\begin{figure}
\centering
\includegraphics[angle=270,scale=0.55]{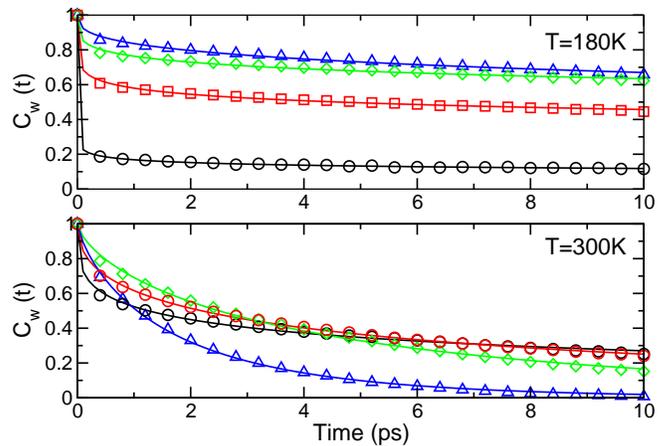}
\caption{Local water polarization correlation function as a function of
time for different depth indices.
The straight lines represent the stretched exponential fits. Symbols are
the same as in Fig. \ref{FIGCa}. Triangles are for bulk water.}
\label{FIGCw}
\end{figure}
As for the backbone, we use stretched exponential fits to the initial 
decay region of the correlation function to extract the relaxation times. 
We depict in Fig. \ref{FIGCw}, for the three typical residues and the bulk, 
the correlation functions along with the corresponding
fits. We note that, here we study the dynamics of a group of water 
molecules, and not the dipolar relaxation of individual water molecules. 
The former permits us to study collectivity and hence evaluate the glassy 
behavior of water clusters around individual residues.

We observe that for both the backbone and the water molecules, the
short time relaxation is relatively well described by the stretched
exponential functions. For longer times a slower, single exponential 
decay also appears in the water correlation functions, attributed to 
the large scale reconfiguration of the water molecules inside the region. 
However, for the current purpose of studying water/backbone dynamics 
on the picosecond time scales, we focus our attention to the former 
relaxations of both the backbone and the hydration shell water.
\begin{figure}
\centering
\includegraphics[angle=270,scale=0.55]{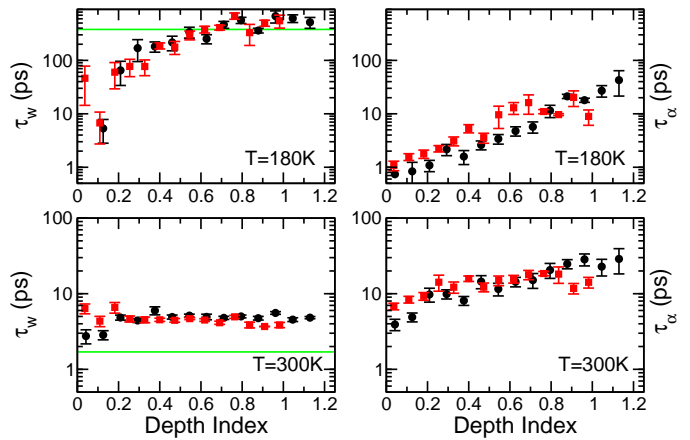}
\caption{Relaxation times of $C_\alpha$ atoms and water molecules as
a function of the depth index for lysozyme (circles) and myoglobin
(squares). The straight lines indicate bulk water relaxation times.} 
\label{FIG4}
\end{figure}
\begin{figure}
\centering
\includegraphics[angle=270,scale=0.55]{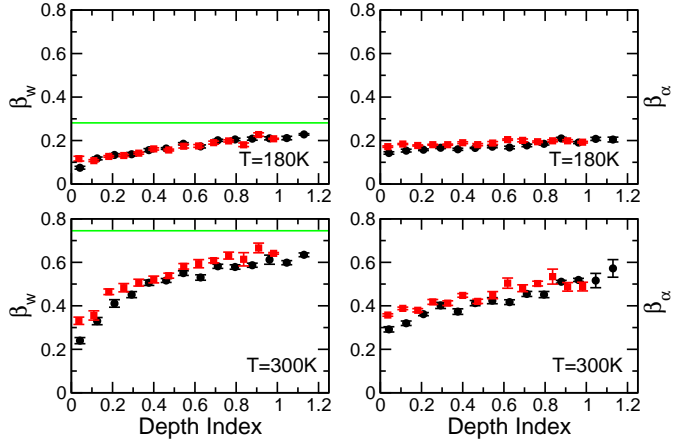}
\caption{Stretched exponents of $C_\alpha$ atoms and
water molecules as a function of the depth index for lysozyme
(circles) and myoglobin (squares). The straight lines indicate bulk
water values.}
\label{FIG4b}
\end{figure}
Having validated the model for the correlation functions, we compute the 
relaxation times for each residue of lysozyme and myoglobin. We depict 
them as a function of depth index in Fig. \ref{FIG4} and their corresponding 
stretched exponents in Fig. \ref{FIG4b} for the two temperatures. We 
observe that the relaxation times of $C_\alpha$ atoms ($\tau_\alpha$) are 
higher at room temperature than at 180 K (the average values are 13 ps 
and 6 ps, respectively). We have previously explained this slowing down 
of the local relaxations using 
a model that relates the onset of the coupling between the local internal 
motions of the protein to the hydration layer with temperature. The results 
showed that such behavior is due to the interplay between the decreased 
stiffness and the modified 
effective friction coefficient. \cite{Atilgan2008} Moreover, independent 
of temperature, $\tau_\alpha$ increases with depth index, which may be 
explained by packing arguments: The more buried a residue is, the larger 
is the coordination number, and hence the local rigidity of the medium. 
Consequently, the magnitude of fluctuations is smaller, leading to faster relaxations.

Conversely, relaxations of water molecules forming the hydration
shell ($\tau_w$) are observed to speed up with increasing
temperature. At 180 K, their values are highly depth index
dependent, accompanied by slowly increasing stretched 
exponents with values close to $\beta = 0.2$. The latter is 
indicative of a collection of trapped states for the rotational motion of 
water molecules according to a model of hierarchically ordered 
materials. \cite{Sornette1997,Shlesinger1986,Jortner1999}
Thus, the water layer shows a strong tendency to be affected by the
local environment (i.e. packing) below $T_d$. However, their
dynamics is independent of depth index at 300 K beyond a critical
depth index ($d$ = 0.3), where there are at least five water
molecules surrounding a given residue (see Fig. \ref{FIG2}), marginally 
enough to form a polarized ``water layer''. The relaxations have a 
characteristic time scale of approximately 5 ps at room temperature. 
The distribution of relaxation times is much narrower as indicated by 
the larger stretched exponents of $\beta > 0.4$ (see Okan {\it et al.} 
\cite{Atilgan2009} for a detailed interpretation of stretched exponents).

The fact that $\tau_w$ are on the same order as $\tau_\alpha$ at 300
K, along with their depth independence suggests that water and
overall protein dynamics are coupled at physiological temperatures,
supporting the idea that water forms a vicinal layer around the
protein, \cite{Atilgan2008} accompanied by a large water
reorganization energy. \cite{Lebard2008} In contrast, $\tau_w$ values 
are $1-2$ orders of magnitude slower than $\tau_\alpha$ at 180 K, along 
with a very strong dependence on the local environment.

Moreover, at room temperature hydration shell water relaxes slower than 
bulk water, compatible with a more viscous layer, slowed down by the 
interactions with the residues. On the other hand, below the transition, 
due to the lack of larger displacement within the 
hydration shell, hydration water relaxes faster than the bulk. This 
suggests that within the time scale of the observations leading to the 
correlation function, the hydration shell water is in a glassy state.

These findings corroborate the results on the residence times in 
Fig. \ref{FIG3}. They lead to a better understanding of the 
relationship between water-protein interactions and the dynamical transition. 
We further investigate this by computing the relaxation times of 
the motions described by Eq. \ref{EQCORRELW} for lysozyme over a wide temperature range.

\subsection{Gradual unfreezing of the hydration layer}
\label{SECTEMPDEP}

\begin{figure}
\centering
\includegraphics[angle=270,scale=0.55]{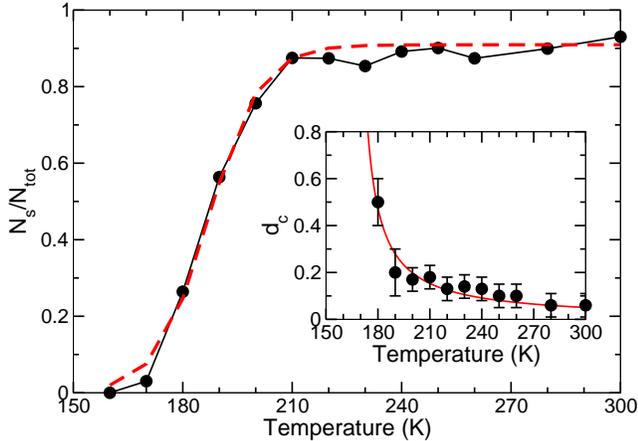}
\caption{Normalized number of residue for which the hydration shell
relaxation time is slower than that of bulk water. The dashed line
represents a Boltzmann sigmoid fit yielding a midpoint value at 187 K. The 
inset represents the critical depth index $d_c$ as a function of temperature, the line is a guide for the eye.}
\label{FIGnsntot}
\end{figure}

We have produced 24 ns trajectories for lysozyme at 13 separate temperatures 
spanning 160 to 300 K and computed the relaxation times of the hydration 
shell and bulk water. We observed in Fig. \ref{FIG4} that below $T_d$, 
hydration shell water relaxes faster than the bulk if the residue's 
location is deep enough. To observe the transition from faster to 
slower relaxation, or in other words from glassy to liquid, one can 
simply count the number of residues slower than the bulk at each temperature 
as depicted in Fig. \ref{FIGnsntot}.
One can then measure the critical depth index $d_C$ beyond which the
hydration shell water becomes slower than the bulk, as shown in the
inset to Fig. \ref{FIGnsntot}. We observe a marked transition from a
state where all the hydration shell waters are faster than the 
bulk to the opposite. The observed behavior may be described by a 
gradual unfreezing of the hydration shell. At temperatures just 
above $T_g$ of water (see the Appendix for a prediction of the $T_g$ 
of the TIP3P water model in the range of 150 - 160 K), only the outermost 
hydration shells are not frozen. For example at 180 K, residues of 
depth index approximately larger than 0.5 have equivalent relaxation 
times to bulk water (also see Fig. \ref{FIG4}). We note 
that some water molecules ($\sim$10$\%$) remain faster than the bulk 
even at temperature well-above $T_d$; these are highly buried water 
molecules that remain coupled to the protein. The gradual 
unfreezing of the hydration shell enhances our understanding of the 
dynamical transition occurring within a wide time and temperature 
range, determined by the time scale of solvent coupled structural 
relaxations. It has already been suggested that the broadening in 
the glass transition of the protein hydration shell is due to a 
distribution of water clusters with different glass transition 
temperatures. \cite{Doster1986} The heterogeneity of the hydration 
shell quantified by the depth index and its relation to the 
hydration levels, residence times and relaxation times expands on this view.

\section{CONCLUSION}
\label{SECCONC}

We study the residence times and memory loss of water
molecules around individual residues to interpret their role in the
dynamics of folded proteins. The calculations are repeated for two
well-studied proteins, leading to identical results: Above $T_d$ the
residence time of water molecules around a given residue may be
interpreted via a simple depth dependent model of trapped diffusion.
The memory loss of the polarization vector displays a similar trend
to that of the protein backbone in terms of the time scales involved.
\cite{Atilgan2009} In contrast, the dynamics of water molecules are
non-diffusive in the regime below $T_d$. Consequently, below $T_d$
the hydration shell is solid-like, behaving as a crust around the
protein. Upon heating, the hydration shell is unfrozen, and permits
the transfer of entropy from the bulk to the protein by the entry/exit 
process of water molecules, hence acting as a plasticizer and 
increasing the overall flexibility of the protein. The contrast 
between the water dynamics in the hydration shell of a protein at 
physiological and non-physiological temperatures therefore reveals 
the interactions and the free energetic requirements necessary to 
achieve biologically relevant motions.

Bulk and hydration layer water have been shown to separately control
motions in functional proteins; for example, the former allows
ligand entry/exit, while the latter is responsible for migration of
ligands within the protein. \cite{Frauenfelder2009} The hydration 
layer acts as a lubricant for the onset of the functional
dynamics in the protein, as new channels between conformational
substates emerge, e.g. through jumps enabled in the side-chain
torsional angles. \cite{Markelz2007} Our results demonstrate 
that hydration water not only performs localized motions, but also 
participates in the global dynamics through long-range diffusion.
Below the transition temperature, on the other hand, the 
miscommunication between the substates maintains the trapped 
hydration water highly dependent on the dynamics of the protein.

\section{APPENDIX}
\subsection{Glass transition temperature of TIP3P water}
\label{SECTIP3PTG}

One of the crucial quantities regarding the dynamical transition in 
proteins is the water glass transition temperature. The glass 
transition temperature is defined as the temperature at which the 
shear viscosity is $10^{12}\; {\rm Pa . s}$, corresponding to a 
relaxation time of approximately 100 s. It is also marked by a 
crossover from ergodic to non-ergodic behavior. Experimentally, 
a step-like decrease in the susceptibilities (e.g. heat capacity) 
is observed. It was suggested that the dynamical transition of the 
protein is dictated by the glass transition of the solvent. 
\cite{Ansari1987,McCammon1989,Karplus2000} Although pure water 
readily forms the ice-phase and is very difficult to vitrify, an 
experimental  value of $\sim 165$ K has nevertheless been measured. 
\cite{Velikov2001,Yue2004} To evaluate the influence of glassy 
water on the protein, we must first assess the behavior of the 
water model used in this study.

\begin{figure}
\centering
\includegraphics[angle=270,scale=0.55]{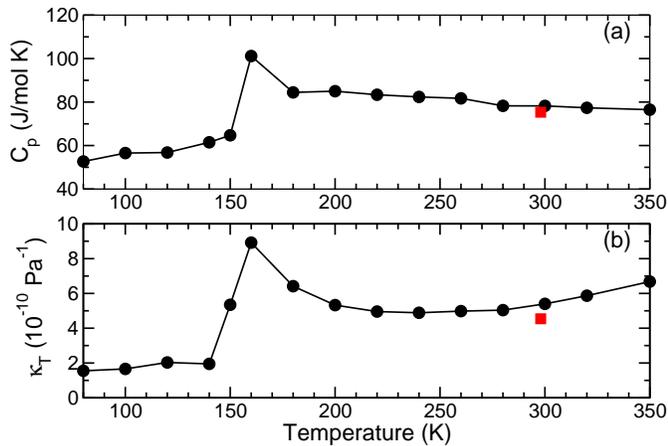}
\caption{Temperature dependence of the (a) isobaric specific heat,
(b) isothermal compressibility. The squares represent the
experimental values at room temperature. \cite{Tipler1999,Halliday1997}} 
\label{FIGCpKp}
\end{figure}

The specific heat may be computed from
\begin{equation}
C=\frac{\langle E^2\rangle-\langle E\rangle^2}{k_B T^2},
\end{equation}
and the isothermal compressibility as
\begin{equation}
\kappa_T=\frac{\langle V^2\rangle-\langle V \rangle^2}
{k_B T \langle V \rangle}\; ,
\end{equation}
where $E$ is the total energy of the system and $V$ the volume. The
results along with experimental values are depicted in Fig. \ref{FIGCpKp}. 
The specific heat and compressibility curves yield a glass transition 
temperature between 150 K and 160 K. This value is lower than the SPC/E, 
TIP4P and TIP5P water models, \cite{Brovchenko2005} though closer to 
the experimental value, \cite{Velikov2001,Yue2004} in accordance with 
the fact that TIP3P water is less structured than the other water models. \cite{Tan2003}
One may also predict the transition by computing the
shear viscosity $\eta$. In equilibrium, one can evaluate $\eta$ 
either from the mean square displacement of the Helfand moment 
associated to viscosity,\cite{Servantie2007a} or from 
the Green-Kubo relation,
\begin{equation}
\eta=\frac{V}{k_BT}\int_0^\infty dt \; \langle P_{\alpha \beta}(t)
P_{\alpha \beta}(0)\rangle,
\label{EQETA}
\end{equation}
where $V$ is the volume of the system, $T$ the temperature, $k_B$ the
Boltzmann constant, and $P_{\alpha \beta}$ the pressure tensor,
\begin{equation}
P_{\alpha \beta}=\frac{1}{V}\left(\sum_im_i v_i^\alpha v_i^\beta+
\sum_{i<j} F_{ij}^\alpha r_{ij}^\beta\right).
\end{equation}
where $\alpha$ and $\beta$ take the values $x$,$y$ and $z$. The
off-diagonal components of $P_{\alpha \beta}$ permit to compute the
shear viscosity $\eta$ while the diagonal ones lead to the bulk
viscosity. Note that the value of the viscosity calculated by 
Eq. \ref {EQETA} is limited by the length of the simulations.
\begin{figure}
\centering
\includegraphics[angle=270,scale=0.55]{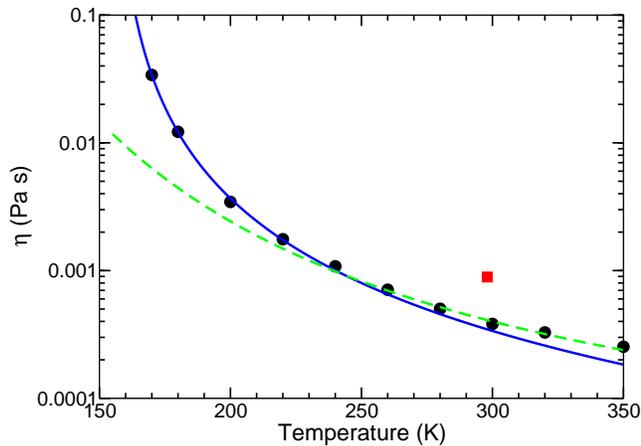}
\caption{Temperature dependence of the shear viscosity $\eta$. The
square represent the experimental value at room temperature. \cite{Franks2000} 
The plain line is the result of the power law fit
at low temperatures and yields $T_g=155$ K and $\gamma=2.1$. The
dashed line is the Arrhenius behavior for high temperatures.}
\label{FIGetaT}
\end{figure}
One can then estimate the glass transition temperature $T_g$ by 
fitting the power law $\eta\sim|T-T_g|^{-\gamma}$. We depict in 
Fig. \ref{FIGetaT} the temperature dependence of the shear viscosity and
the power law fit. The fit yields a glass transition temperature
$T_g=155$ K in accord with the values obtained from the susceptibilities.

\section*{ACKNOWLEDGMENTS}
This research is supported by the DSAP grant of the Turkish Academy
of Sciences (T\"UBA), and is partially funded by the Scientific and
Technological Research Council of Turkey Project No. 106T522.


\end{document}